\documentclass[11pt,a4paper]{article}
\pdfoutput=1

\usepackage{jheppub}
\usepackage{amsfonts,amsmath,bm}
\usepackage{array}
\usepackage{graphicx}
\usepackage{caption}
\makeatletter
\def\@fpheader{\relax}
\makeatother

\newcommand\be{\begin{equation}}
\newcommand\ee{\end{equation}}
\newcommand\bea{\begin{eqnarray}}
\newcommand\eea{\end{eqnarray}}
\newcommand\eref[1]{(\ref{#1})}

\newcommand\capt[1]{\caption{\textsf{#1}}}
\newcommand\comment[1]{}

\title{\centering A Tale of Two Compact Bosons}

\author[*]{Christian Ferko,}
\author[\dagger,*]{Vishnu Jejjala,}
\author[\ddagger]{Brandon Robinson}

\affiliation[*]{The NSF Institute for Artificial Intelligence and Fundamental Interactions (IAIFI)\\
and Department of Physics, Northeastern University, Boston, MA 02115, USA}

\affiliation[\dagger]{Mandelstam Institute for Theoretical Physics, School of Physics, and NITheCS,\\
University of the Witwatersrand, Johannesburg, WITS 2050, South Africa}

\affiliation[\ddagger]{Institute of Physics, University of Amsterdam, Science Park 904,\\ 1098 XH Amsterdam, Netherlands}

\emailAdd{c.ferko@northeastern.edu}
\emailAdd{v.jejjala@wits.ac.za}
\emailAdd{b.j.robinson@uva.nl}

\abstract{%
Neural network field theory (NN-FT) defines a field theory by a network architecture together with a probability density on its latent variables. For compact theories the local Gaussian sector is only part of the story: one must also sum over discrete topological sectors. We review a mixed continuous/discrete latent-variable construction and apply it to two compact bosons. For the Berezinskii--Kosterlitz--Thouless transition, a random Fourier feature spin-wave sampler supplemented by an explicit Coulomb gas vortex sector reproduces the Gaussian critical line below $T_c$, vortex proliferation above $T_c$, the essential singularity of the correlation length, and the Nelson--Kosterlitz jump. For the bosonic string, oscillator modes augmented by momentum--winding labels reproduce circle T-duality, Buscher transformations on constant toroidal backgrounds, self-dual current algebra enhancement, and a toy T-fold. The common lesson is that the same local neural sampler, paired with different discrete topological data, yields physically distinct compact theories. These proceedings are based on~\cite{Ferko:2026ken}.
}

\begin{document}
\notoc
\compress
\maketitle
\newpage

\section{Introduction}

Neural network field theory (NN-FT) replaces the usual sum over fields by a statistical ensemble over network parameters~\cite{Halverson:2020trp,Halverson:2021aot}. In the large width limit many architectures become Gaussian processes, so free fields arise directly from the central limit theorem; finite-width effects or non-Gaussian parameter densities then generate interactions~\cite{Demirtas:2023fir,Ferko:2026axm}. This viewpoint has already been applied to symmetry realization, conformal field theory, and string worldsheet models~\cite{Maiti:2021fpy,Halverson:2024axc,Robinson:2025ybg,Frank:2026bui}.

The question addressed here is how to incorporate topology. A non-compact Gaussian covering field captures smooth local fluctuations, but a compact boson also carries discrete data from the start: vortex sectors in the two-dimensional $XY$ model, or momentum and winding sectors in the bosonic string. In an ordinary path integral such sectors enter through the field space and measure, not through a Gaussian fluctuation integral alone. NN-FT should therefore be enlarged in the same spirit.

The basic proposal is to work with mixed latent variables
\be
\Theta=(\theta,Q)~,
\ee
where $\theta$ denotes continuous neural parameters and $Q$ labels discrete topological sectors. Observables are then computed as
\be
\langle \mathcal O\rangle = \sum_Q \int d\theta\; P(\theta,Q)\,\mathcal O[\phi_{\theta,Q}]~.
\label{eq:mixed}
\ee
The continuous variables describe the local Gaussian sector that a random feature network already knows how to sample, while the discrete labels keep track of global data such as vortices, momentum, or winding.

We discuss two case studies. The first is the Berezinskii--Kosterlitz--Thouless (BKT) transition, where the relevant topology is dynamical: bound vortex--antivortex pairs at low temperature unbind above a critical point~\cite{Berezinskii1971,KosterlitzThouless1973,Kosterlitz1974,Jose1977}. The second is T-duality of the bosonic string, where topology is kinematical: the same compact conformal field theory admits dual descriptions related by exchange of momentum and winding~\cite{Buscher:1987sk,Buscher:1987qj}. In both examples the local Gaussian sector is essentially the same, but distinct discrete topological data lead to distinct physics.

A more complete treatment can be found in~\cite{Ferko:2026ken}.

\section{Neural network field theory and compact sectors}

For a scalar architecture $\phi_\theta(x)$ on Euclidean spacetime, NN-FT defines correlators by parameter space averages,
\be
G_n(x_1,\ldots,x_n)=\int d\theta\;P(\theta)\,\phi_\theta(x_1)\cdots\phi_\theta(x_n)~.
\label{eq:gn}
\ee
Equivalently, one may write the generating functional as
\be
Z[J]=\int d\theta\;P(\theta)\,\exp\!\left(\int d^dx\,J(x)\phi_\theta(x)\right)~.
\label{eq:ZJ}
\ee
When the induced Schwinger functions satisfy the Osterwalder--Schrader axioms, this defines a bona fide Euclidean quantum field theory~\cite{Osterwalder:1973dx,Osterwalder:1974tc,Halverson:2021aot}.

A simple large width architecture is the random Fourier feature expansion
\be
\phi_N(x)=\frac{1}{\sqrt N}\sum_{i=1}^N a_i\cos(w_i\!\cdot\! x+b_i)~,
\label{eq:rff}
\ee
with Gaussian amplitudes $a_i$, random phases $b_i$, and frequencies $w_i$ sampled from a spectral density $\rho(w)$~\cite{rahimi2007random}. As $N\to\infty$, the field becomes Gaussian with two-point kernel
\be
K(x,x')=\frac{\sigma_a^2}{2}\int d^dw\,\rho(w)\cos\big(w\!\cdot\!(x-x')\big)~.
\label{eq:kernel}
\ee
This realizes the local free sector of many field theories~\cite{Halverson:2020trp,Halverson:2021aot,Frank:2026bui}. What it does \emph{not} do by itself is generate topological sectors of a compact target.

The compact boson is the cleanest example. A real Gaussian covering field lives on $\mathbb R$ and is globally single valued, whereas a compact boson lives on $S^1$ and admits winding, momentum, vortices, and defects. Equation~\eref{eq:mixed} is therefore not an optional embellishment but part of the definition of the compact theory. This point can be summarized schematically as
\be
\text{same local Gaussian sector} + \text{different topological labels}
\quad \Longrightarrow \quad \text{different physics}~.
\label{eq:slogan}
\ee
The remainder of the paper makes this slogan concrete.

\section{First compact boson: BKT and the dynamics of topology}

The Mermin--Wagner theorem forbids spontaneous breaking of continuous symmetries in two spatial dimensions with sufficiently short range interactions~\cite{MerminWagner1966}. Nevertheless, the two-dimensional $XY$ model has a low-temperature phase with quasi-long-range order and a topological phase transition driven by vortex unbinding~\cite{Berezinskii1971,KosterlitzThouless1973,Kosterlitz1974,NelsonKosterlitz1977}. In the infrared the model flows to a compact boson,
\be
S_\text{sw}=\frac{K_R}{2}\int d^2x\,(\nabla\theta)^2~,
\qquad \theta(x)\sim\theta(x)+2\pi~,
\label{eq:swaction}
\ee
with dimensionless stiffness $K_R$. Vertex operators $V_q(x)=e^{iq\theta(x)}$ have correlators
\be
G_q(x)=\langle V_q(x)V_{-q}(0)\rangle
\sim
\begin{cases}
|x|^{-q^2/(2\pi K_R)}~, & T<T_c~,\\[3pt]
e^{-|x|/\xi(T)}~, & T>T_c~,
\end{cases}
\label{eq:bktcorr}
\ee
where the high-temperature phase has finite correlation length with the characteristic BKT essential singularity.

The NN-FT realization separates the smooth and topological sectors. The spin-wave field is sampled by a random Fourier feature architecture on an $L\times L$ torus,
\be
\theta_\text{sw}(x)=\frac{A}{\sqrt N}\sum_{j=1}^N \cos(k_j\!\cdot\!x+\gamma_j)~,
\qquad k_j=\frac{2\pi}{L}n_j~, 
\label{eq:thetann}
\ee
with integer modes $n_j\in\mathbb Z^2$ drawn from $p(n)\propto |n|^{-2}$ and phases $\gamma_j$ uniform on $[0,2\pi]$. The compact spin field is the projection to $S^1$,
\be
\bm s(x)=\big(\cos(b\,\theta_\text{sw}(x)),\,\sin(b\,\theta_\text{sw}(x))\big)~,
\label{eq:spinfield}
\ee
or equivalently the vertex operator $V_b(x)=e^{ib\theta_\text{sw}(x)}$. In the spin-wave sector one finds
\be
\langle V_b(x)V_b^*(0)\rangle\sim |x|^{-b^2}~,
\qquad \eta_\text{sw}=b^2~,
\label{eq:eta}
\ee
so matching to the compact boson gives
\be
K_0=\frac{1}{2\pi b^2}~,
\qquad b_c=\frac12~,
\qquad T_c=\frac{\pi}{2}
\label{eq:dictionary}
\ee
in units where the bare stiffness is set to one.

The vortex sector is added explicitly through a neutral Coulomb gas,
\be
\theta(x)=b\,\theta_\text{sw}(x)+\theta_v(x)~,
\qquad
\theta_v(x)=\sum_{a=1}^{N_v}m_a\,\arg(x-x_a)~,
\label{eq:vortexfield}
\ee
with charges $m_a=\pm 1$ satisfying $\sum_a m_a=0$, and probability distribution
\be
P_\text{vort}(\mathcal V)\propto y^{N_v}
\exp\!\left[2\pi K_0\sum_{a<b}m_am_b\log(r_{ab}+a_c)\right]~.
\label{eq:cvort}
\ee
Because the spin-wave and vortex sectors are sampled independently, the full two-point function factorizes,
\be
G_2(r)=\langle\cos(\theta(x)-\theta(0))\rangle=G_\text{sw}(r)\,G_v(r)~.
\label{eq:factor}
\ee
At low temperature, $G_v(r)$ tends to a constant because vortices are tightly bound into neutral pairs. At high temperature, free vortices proliferate and drive the correlator to exponential decay.

\begin{figure}[t]
\centering
\includegraphics[width=0.62\textwidth]{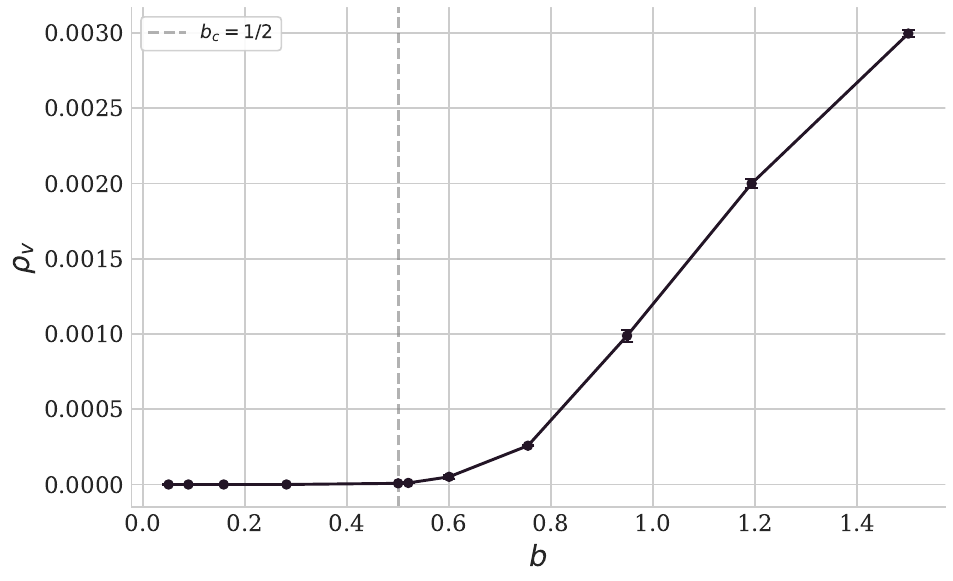}
\caption{Vortex density $\rho_v$ across the BKT transition in the mixed NN-FT ensemble. The density is essentially zero below the critical point $b_c=1/2$ and rises rapidly above it, signaling the onset of free vortices.}
\label{fig:vortex_density_proc}
\end{figure}

This construction reproduces the qualitative BKT phenomenology. The spin-wave correlator follows the critical line $\eta=b^2$ with high precision. Once the Coulomb gas sector is turned on, the vortex density remains negligible for $b<b_c$ and rises rapidly for $b>b_c$, giving a direct numerical signature of vortex proliferation; this is shown in Figure~\ref{fig:vortex_density_proc}. The pair correlation $g_{+-}(r)$ likewise evolves from a strongly bound regime to one approaching the uncorrelated value $1$, signaling unbinding. Above the transition the full correlator is well fit by
\be
G_2(r)\propto r^{-b^2}e^{-r/\xi}~,
\qquad
\xi(b)\sim \exp\!\left[\frac{c}{\sqrt{b^2-b_c^2}}\right]~,
\label{eq:xi}
\ee
in agreement with the BKT essential singularity.

\begin{figure}[t]
\centering
\includegraphics[width=0.72\textwidth]{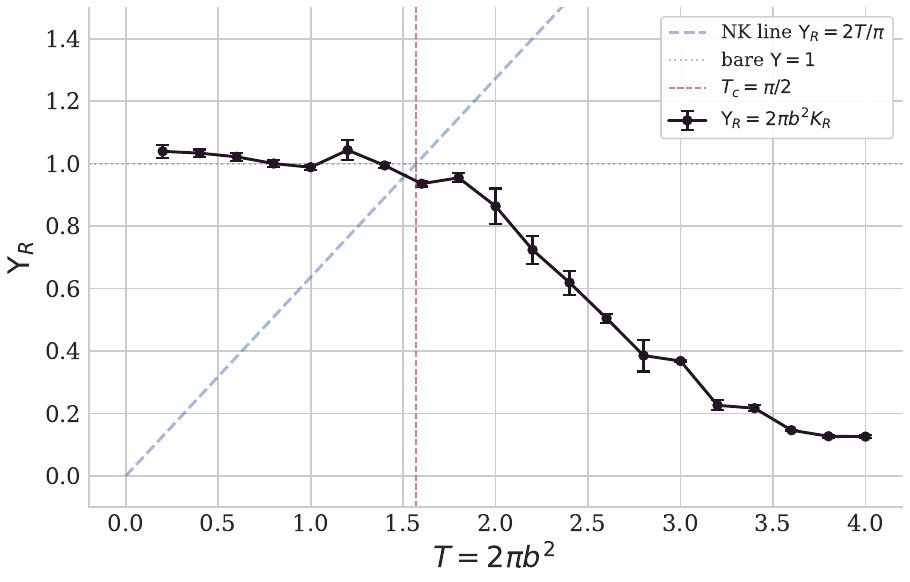}
\caption{Helicity modulus $\Upsilon_R=2\pi b^2 K_R$ as a function of temperature $T=2\pi b^2$. The data cross the Nelson--Kosterlitz line near $T_c=\pi/2$ and collapse in the disordered phase, in accord with the expected universal jump.}
\label{fig:helicity_proc}
\end{figure}

The renormalized stiffness extracted from the full field defines a helicity modulus
\be
\Upsilon_R = T K_R~,
\label{eq:helicity}
\ee
which crosses the Nelson--Kosterlitz line near $T_c=\pi/2$ and then drops toward zero in the disordered phase~\cite{NelsonKosterlitz1977}. Figure~\ref{fig:helicity_proc} shows this characteristic BKT behavior in the NN-FT data. Without the discrete vortex sector none of this happens: the theory remains on the Gaussian critical line for all $b$. The BKT transition therefore furnishes an example in which topology is not optional bookkeeping but a dynamical ingredient of the NN-FT.

\section{Second compact boson: T-duality and the kinematics of topology}

For the bosonic string, compactness manifests itself through momentum and winding sectors rather than through thermally excited vortices. The Euclidean worldsheet sigma model is
\be
\begin{aligned}
S[X;h]=\frac{1}{4\pi\alpha'}\int d^2\sigma\,\sqrt h\,
\Big[&h^{ab}G_{\mu\nu}(X)\partial_aX^\mu\partial_bX^\nu\\
&+i\epsilon^{ab}B_{\mu\nu}(X)\partial_aX^\mu\partial_bX^\nu
+\alpha'\Phi(X)R^{(2)}[h]\Big]~.
\end{aligned}
\label{eq:sigma}
\ee
When the target has a $U(1)$ isometry, T-duality relates one background to another with radius $R\leftrightarrow \widetilde R=\alpha'/R$ and with sigma model couplings transformed by the Buscher rules~\cite{Buscher:1987sk,Buscher:1987qj}. For a circle,
\be
X(\tau,\sigma)=X_\text{osc}(\tau,\sigma)+x_0+\frac{\alpha' n}{R}\,\tau+wR\,\sigma~,
\label{eq:circlemode}
\ee
so that
\be
p_L=\frac{n}{R}+\frac{wR}{\alpha'}~,
\qquad
p_R=\frac{n}{R}-\frac{wR}{\alpha'}~,
\label{eq:plpr}
\ee
and under T-duality
\be
R\longrightarrow \widetilde R=\frac{\alpha'}{R}~,
\qquad (n,w)\longrightarrow (w,n)~,
\qquad p_L\longrightarrow p_L~,
\qquad p_R\longrightarrow -p_R~.
\label{eq:tduality}
\ee

The NN-FT realization again has mixed latent variables. The oscillator sector is sampled by a Gaussian neural architecture that reproduces the free boson propagator on the worldsheet~\cite{Frank:2026bui}; the compact sector is labeled by discrete momentum and winding. For any observable,
\be
\langle\mathcal O\rangle_R=
\sum_{n,w\in\mathbb Z}\int d\Theta_\text{osc}\;P_\text{osc}(\Theta_\text{osc})P_R(n,w)
\mathcal O[X_{\Theta_\text{osc};n,w}]~,
\label{eq:compactmix}
\ee
with lattice weights
\be
P_R(n,w)\propto
\exp\!\left[-\pi\tau_2\left(\frac{\alpha'n^2}{R^2}+\frac{R^2w^2}{\alpha'}\right)\right]~.
\label{eq:weights}
\ee
This mixed ensemble reproduces several layers of the usual duality.

First, the circle duality is realized sample by sample. The discrete weights satisfy the exact truncated lattice identity $P_R(n,w)=P_{\widetilde R}(w,n)$ to machine precision, and the sampled monodromy
\be
\Delta_\sigma X = X(\tau,2\pi)-X(\tau,0)=2\pi wR
\label{eq:monodromy}
\ee
transforms exactly into its dual counterpart with exchanged momentum and winding. The zero mode slopes are exchanged as well.

Second, on constant two-torus backgrounds the same construction reproduces the Buscher transformation of the background fields themselves. Writing the isometry direction as $y$, the relevant part of the transformation is
\be
\widetilde G_{yy}=\frac{1}{G_{yy}}~,
\qquad
\widetilde G_{yi}=\frac{B_{yi}}{G_{yy}}~,
\qquad
\widetilde B_{yi}=\frac{G_{yi}}{G_{yy}}~,
\qquad
\widetilde\Phi=\Phi-\frac12\log G_{yy}~,
\label{eq:buscher}
\ee
with analogous expressions for $\widetilde G_{ij}$ and $\widetilde B_{ij}$~\cite{Buscher:1987sk,Buscher:1987qj}. In the NN-FT, the oscillator covariance tracks the inverse metric, while the compact lattice carries the $G$- and $B$-dependence of the zero modes. Numerically one finds that the transformed lattice weights, monodromies, and marginals agree with the Buscher map, and that the lower-dimensional invariant $e^{-2\Phi}\sqrt{\det G}$ is preserved.

Third, operator data are correctly organized. With explicit chiral fields $X_L$ and $X_R$, the compact boson vertex operators
\be
V_{n,w}(u,v) =\ :\!\exp\!\big(i p_L X_L(u)+i p_R X_R(v)\big)\!:\,
\label{eq:vertex}
\ee
have conformal weights $(h,\bar h)=(\alpha' p_L^2/4,\alpha' p_R^2/4)$. Zero mode averaging enforces the usual neutrality rules, while paired draws at radius $R$ and at $\widetilde R$ show that momentum operators at one radius match winding operators at the dual radius. At the self-dual point $R=\sqrt{\alpha'}$, the NN-FT does more than preserve the spectrum: it reproduces the enhancement of $U(1)_L\times U(1)_R$ to $SU(2)_L\times SU(2)_R$, with charged currents sitting at dimension one.

Finally, the same language also accommodates a simple non-geometric example. Patchwise torus samplers can be glued together by a Buscher transformation so that each local chart is geometric, but global closure is achieved only after an $O(2,2;\mathbb Z)$ action. In other words, the NN-FT can represent a toy T-fold~\cite{Hull:2004in,Hull:2006qs,Hull:2006va}.

\section{Example: the quantum rotor}
The moral can be stated in the simplest arena: compactness is not a property of a local Gaussian sampler alone. It is a statement about the field space. The quantum rotor, or compact boson in Euclidean quantum mechanics, has
\be
S[\varphi]=\frac{I}{2}\int_0^\beta d\tau\,\dot\varphi^2~,\qquad
\varphi(\tau)\sim\varphi(\tau)+2\pi~,
\label{eq:rotor_action}
\ee
and provides a one-dimensional test of the mixed NN-FT ensemble. This is the compact analogue of the Euclidean neural network quantum mechanics constructions of~\cite{Ferko:2025ogz}.

The architecture separates a continuous oscillator sector from an integer winding sector,
\be
\varphi_{\theta,m}(\tau)=\varphi_0+\frac{2\pi w}{\beta}\tau+\eta_{\theta}(\tau)~,\qquad
w\in\mathbb Z~.
\label{eq:rotor_architecture}
\ee
Here $\varphi_0$ is uniformly distributed on $[0, 2 \pi)$, and with a mode cutoff $M$,
\be
\eta_{\theta}(\tau)=\sum_{\ell=1}^{M}\left[a_\ell\cos\frac{2\pi\ell\tau}{\beta}+b_\ell\sin\frac{2\pi\ell\tau}{\beta}\right]~,\qquad
a_\ell,b_\ell\sim {\cal N}\!\left(0,\sigma_\ell^2\right)~,\quad
\sigma_\ell^2=\frac{\beta}{2\pi^2 I\ell^2}~.
\label{eq:rotor_modes}
\ee
The winding action is $S_w=2\pi^2 I w^2/\beta$, so
\be
\langle{\cal O}\rangle=\sum_{w\in\mathbb Z}\int d\theta\;P_\text{osc}(\theta)\, P_\beta(w)\,{\cal O}[\varphi_{\theta,w}]~,
\qquad
P_\beta(w)\propto e^{-2\pi^2 I w^2/\beta}~.
\label{eq:rotor_mixed}
\ee
The parameter space integrals are elementary. For $q\in\mathbb Z$, define the vertex correlator $C_q(\tau)=\langle e^{iq\Delta\varphi}\rangle$, where $\Delta\varphi=\varphi(\tau)-\varphi(0)$. With $c_\ell=\cos(2\pi\ell\tau/\beta)$ and $s_\ell=\sin(2\pi\ell\tau/\beta)$,
\be
\begin{aligned}
C_q^{(M)}(\tau)
&= \frac{\sum_w e^{-2\pi^2 I w^2/\beta}e^{2\pi iqw\tau/\beta}}{\sum_w e^{-2\pi^2 I w^2/\beta}} \prod_{\ell=1}^{M} \int\frac{da_\ell db_\ell}{2\pi\sigma_\ell^2} e^{-\frac{a_\ell^2+b_\ell^2}{2\sigma_\ell^2}} e^{iq[a_\ell(c_\ell-1)+b_\ell s_\ell]} \\
&= \exp\!\left[-\frac{q^2}{2}V_M(\tau)\right]\, \frac{\sum_w e^{-2\pi^2 I w^2/\beta}e^{2\pi iqw\tau/\beta}} {\sum_w e^{-2\pi^2 I w^2/\beta}}~,\qquad
V_M(\tau)=2\sum_{\ell=1}^M\sigma_\ell^2(1-c_\ell)~.
\end{aligned}
\label{eq:rotor_integral}
\ee
As $M\to\infty$, $V_M(\tau)\to \tau(\beta-\tau)/(I\beta)$, giving
\be
C_q(\tau) = \exp\!\left[-\frac{q^2\tau(\beta-\tau)}{2I\beta}\right] \frac{\vartheta_3\!\left(\pi q\tau/\beta,e^{-2\pi^2 I/\beta}\right)} {\vartheta_3\!\left(0,e^{-2\pi^2 I/\beta}\right)}~.
\label{eq:rotor_winding_correlator}
\ee
The theory can be solved exactly.
The partition function has two Poisson dual expressions:
\be
Z(\beta) = \sqrt{\frac{2\pi I}{\beta}}\sum_{w\in\mathbb Z}e^{-2\pi^2 I w^2/\beta} =\sum_{n\in\mathbb Z}e^{-\beta n^2/(2I)}~,
\ee
and the exact thermal correlator is
\be
C_q(\tau) = {1\over Z}\sum_{n\in\mathbb Z} \exp\!\left[-{(\beta-\tau)n^2\over 2I}-{\tau(n+q)^2\over 2I}\right]~.\\
\label{eq:rotor_poisson}
\ee
Using Poisson resummation and the definition of the Jacobi theta function $\vartheta_3$,~\eref{eq:rotor_winding_correlator} and~\eref{eq:rotor_poisson} are identical. The mixed continuous/discrete neural network quantum mechanics (NN-QM) construction, with winding sectors, is exactly equivalent to the conventional Hamiltonian formulation of the quantum rotor; $e^{iq\varphi}$ shifts angular momentum as $n\mapsto n+q$. As a final check, Table~\ref{tab:rotor} summarizes the comparison between exact results versus the Monte Carlo sampling for the numerical implementation and confirms that the architecture reproduces the expected physics.

\begin{table}[t]
\centering
\small
\begin{tabular}{c c c c c}
\hline
$\tau/\beta$ & exact $C_1$ & mixed MC & MC s.e. & $|\Delta|$ \\
\hline
0.10 & 0.753267 & 0.755627 & $6.8\times10^{-4}$ & $2.4\times10^{-3}$ \\
0.25 & 0.530265 & 0.531714 & $1.1\times10^{-3}$ & $1.4\times10^{-3}$ \\
0.50 & 0.406850 & 0.406679 & $1.3\times10^{-3}$ & $1.7\times10^{-4}$ \\
0.75 & 0.530265 & 0.529827 & $1.1\times10^{-3}$ & $4.4\times10^{-4}$ \\
0.90 & 0.753267 & 0.753166 & $6.9\times10^{-4}$ & $1.0\times10^{-4}$ \\
\hline
\end{tabular}
\capt{Quantum rotor check for $\beta=6$, $I=1$, $q=1$, $M=256$, and $2\times10^5$ Monte Carlo samples. The code samples $w$ and the finite mode Gaussian marginal implied by~\eref{eq:rotor_modes}. Columns show $\tau/\beta$, the exact correlator $C_1(\tau)$, the mixed ensemble Monte Carlo estimate, its standard error, and the absolute error $|\Delta|=|C_1^\text{MC}-C_1^\text{exact}|$.}
\label{tab:rotor}
\end{table}

\section{Discussion}

The compact bosons studied here illustrate complementary roles for topology in NN-FT. In the BKT transition, topology is dynamical: the vortex sector changes across the transition and converts algebraic order into exponential decay. In T-duality, topology is kinematical: different momentum--winding labels describe dual presentations of the same compact conformal field theory. In both cases the neural sampler for local Gaussian fluctuations is straightforward, but the full physics only appears once discrete topological data are included explicitly.

The rotor in NN-QM is a minimal example in the same spirit as the compact bosons we have seen in NN-FT. The continuous variables $\theta$ produce the inverse Laplacian Gaussian fluctuations, but the compact theory is obtained only after summing over $w\in\mathbb Z$. As with the BKT vortices and string momentum--winding, the quantum rotor evinces sectors that exhibit different topological data. The essential lesson is the same: the correct ensemble is mixed continuous/discrete.

The resulting picture is constructive. Rather than asking an unconstrained network to discover topology automatically, one enlarges parameter space so that smooth fluctuations and topological sectors are sampled together. This mirrors the ordinary path integral treatment of compact theories and suggests a general template for gauge theories, theories with defects, non-trivial bundles, and non-geometric backgrounds. Mixed continuous/discrete latent variables therefore provide a natural route by which NN-FT can move beyond local Gaussian physics toward theories with genuine global structure.

\acknowledgments
We are grateful to Jim Halverson for collaboration on the paper~\cite{Ferko:2026ken} on which these proceedings are based, for discussions about NN-FT in this and other contexts, and for extensive, illuminating conversations about Physics in the Age of LLMs.
We thank Miranda Cheng, Sergei Gukov, Elli Heyes, and Ed Hirst for organizing the workshop ``DANGER: Data, Numbers, and Geometry'' at the Banff International Research Station from 5-10 April 2026 and the participants for a fascinating meeting.
C.F.\ is supported by the National Science Foundation under Cooperative Agreement PHY-2019786.
V.J.\ is supported by the South African Research Chairs Initiative of the Department of Science, Technology, and Innovation and the National Research Foundation (grant 78554).
B.R.\ is supported by NWO vidi grant 016.Vidi.189.182.
The initial draft of the proceedings were generated by GPT-5.5 from~\cite{Ferko:2026ken} and a workshop talk by one of us (V.J.)~\cite{Jejjala:2026Banff} and edited by the authors.

\bibliographystyle{JHEP}
\bibliography{refs}

\end{document}